\documentclass{llncs}

\newif\iffullversion

\fullversiontrue

\usepackage[lineseq,shortlogic,shortbib]{my}
\usepackage{amsmath,cite,url,listings}
\usepackage{tikz}
\usetikzlibrary{shapes.geometric, arrows}

\def\RR{{\mathcal{R}}}
\def\UU{{\mathcal{U}}}
\def\PP{{\mathcal{P}}}

\def\Amax{{\mathcal{M}ax}}
\def\Ams{{\mathcal{MS}um}}
\def\Apol{{\mathcal{P}\kern-.1emol}}
\def\Amp{{\mathcal{M}\Apol}}

\def\VGS{\succsim}
\def\VGT{\succ}
\def\VGSopt{\mathrel{\SuperImpose\succsim{\lower.5ex\hbox{\rm\tiny(\hskip 1.1em)}}}}

\def\SCC{\mathit{SCC}}

\title{Nagoya Termination Tool%
\iffullversion
	\thanks{Full version of the paper which is to appear in the
		\textit{Proceedings of the Joint
		25th International Conference on Rewriting Techniques and Applications
		and 12th International Conference on Typed Lambda Calculi and Applications
		(RTA-TLCA\;'14)},
		LNCS Advanced Research in Computing and Software Science,
		Springer, 2014.
	}%
\fi
}
\titlerunning{NaTT}
\author{Akihisa Yamada\inst{1} \and Keiichirou Kusakari\inst{2} \and Toshiki Sakabe\inst{1}}
\authorrunning{A\@. Yamada, K\@. Kusakari, T\@. Sakabe}

\institute{
	Graduate School of Information Science, Nagoya University, Japan
\and
	Faculty of Engineering, Gifu University, Japan
}

\begin{document}

\maketitle

\begin{abstract}
This paper describes the implementation and techniques of
the Nagoya Termination Tool,
a termination \REV{prover}{analyzer} for term rewrite systems.
\REV{%
	The main features of the tool are:
	the first implementation of
}{%
	The tool is special for its power due to
	the implementation of
}%
the weighted path order
which subsumes most of the existing reduction pairs,
and
the efficiency due to the
\REV{strong cooperation with external}{deep cooperation with}
SMT solvers.
We present some new ideas that contribute to 
the efficiency and \REV{}{the }power of the tool.
\end{abstract}

\section{Introduction}

Proving termination of term rewrite systems (TRSs) has been an
active field of research.
In this paper, we describe the \emph{Nagoya Termination Tool} (\NaTT),
a termination prover for TRS, which is available at
\begin{center}
\url{http://www.trs.cm.is.nagoya-u.ac.jp/NaTT/}
\end{center}

\NaTT \REV{is powerful and fast; its}{has the advantages of power and efficiency. Its}
power comes from the novel implementation of the
\emph{weighted path order (WPO)} \cite{YKS13,YKS14} that
subsumes most of the existing reduction pairs, and
its efficiency comes from the \REV{strong}{deep} cooperation with
state-of-the-art \emph{satisfiability modulo theory (SMT)} solvers.
In principle, any solver that complies with
the \SMTLIB Standard\footnote{\url{http://www.smtlib.org/}}
version 2.0 can be incorporated
as a back-end \REV{into}{of} \NaTT.

In the next section,
we recall the dependency pair framework that \NaTT is based on,
and present existing techniques that are implemented in \NaTT.
Section~\ref{sec:reduction} describes the implementation of WPO and
demonstrates how to obtain other existing techniques as instances of WPO.
Some techniques on cooperating with SMT solvers are presented in Section~\ref{sec:SMT}.
After giving some design details in Section~\ref{sec:design},
we assess the tool by \REV{its results in}{the result of} the \emph{termination competition}%
\footnote{\url{http://termination-portal.org/wiki/Termination_Competition}}
in Section~\ref{sec:assessment}.
Then we conclude in Section~\ref{sec:conclusion}.
\REV{%
}{}%

\section{The Dependency Pair Framework}

The overall procedure of \NaTT is illustrated in Figure~\ref{fig:flow}.
\begin{figure}[tb]
\tikzstyle{process} = [rectangle, minimum width=3cm, minimum height=16pt, text centered, draw=black]%
\tikzstyle{decision} = [diamond, minimum width=40pt, minimum height=16pt, text centered, draw=black, aspect=3]%
\tikzstyle{arrow} = [thick,->,>=stealth]%
\centering
\fbox{
\begin{tikzpicture}[node distance=30pt]
\node(start) at (0,-.1){Input TRS};
\node(remove)[process, below of=start]{Rule Removal Processors};
\node(removed)[decision, below of=remove, yshift=-5pt]{Success?};
\node(uncurry)[process, below of=removed, yshift=-5pt]{Uncurrying Processor};
\node(EDG)[process, below of=uncurry]{EDG Processor};
\node(test)[decision] at (5.5,-.8) {$\SCC = \emptyset$?};
\node(yes) at (test.east) [xshift=40pt]{\textsf{yes}};
\node(reduce)[process, below of=test,yshift=-5pt]{Reduction Pair Processors};
\node(reduced)[decision, below of=reduce]{Success?};
\node(looping)[decision, below of=reduced, yshift=-10pt]{Find loop?};
\node(no) at (looping.east) [xshift=40pt]{\textsf{no}};
\node(maybe) at (no.south) [yshift=-20pt, xshift=-10pt]{\textsf{maybe}};
\draw[arrow] (start) -- (remove);
\draw[arrow] (remove) -- (removed);
\draw[arrow] (removed) -- node[anchor=west]{no} (uncurry);
\draw[arrow] (removed) -| node[anchor=north west]{yes} ++(-2.5,0) |- (remove);
\draw[arrow] (uncurry) -- (EDG);
\draw[arrow] (EDG) -| (2.5,0) -| (test);
\draw[arrow] (test) -- node[anchor=south]{yes} (yes);
\draw[arrow] (test) -- node[anchor=west]{no} (reduce);
\draw[arrow] (reduce) -- (reduced);
\draw[arrow] (reduced) -| node[anchor=north,xshift=10pt]{yes} ++(-2.5,0) |- (test);
\draw[arrow] (reduced) -- node[anchor=west]{no} (looping);
\draw[arrow] (looping) -- node[anchor=south]{found} (no);
\draw[arrow] (looping) |- node[anchor=south west]{not found} (maybe);
\end{tikzpicture}}
\caption{Flowchart of \NaTT\label{fig:flow}}
\end{figure}
\NaTT is based on 
the \emph{dependency pair framework (DP framework)} \cite{AG00,HM04,GTSF06},
a very successful technique for proving termination of TRSs
which is implemented in almost
all the modern termination provers for TRSs.
In the DP framework, 
\REV{%
	dependencies between
}{%
	dependency of
}%
function calls defined in a TRS $\RR$ is
expressed by the set $\DP(\RR)$ of \emph{dependency pairs}.
\REV{%
	If a function $f$ is defined by a rule 
	\[
		f(s_1,\dots,s_n) \to C[g(t_1,\dots,t_m)] \in \RR
	\]
	where $g$ is also defined in $\RR$, then
	this dependency is described by the following dependency pair:
}{%
	A dependency pair, written as a rewrite rule\REV{}{:}
}%
	\[
		f^\sharp(s_1,\dots,s_n) \to g^\sharp(t_1,\dots,t_m) \in \DP(\RR)
	\]
\REV{}{%
	\[
		f^\sharp(s_1,\dots,s_n) \to g^\sharp(t_1,\dots,t_m)\ \in\ \DP(\RR)
	\]
	describes that the function $f$ is defined by a rule 
	\[
		f(s_1,\dots,s_n) \to C[g(t_1,\dots,t_m)]\ \in\ \RR
	\]
	and $g$ is also defined in $\RR$.
}%
The DP framework (dis)proves termination of $\RR$ by
simplifying and decomposing \emph{DP problems} $\Tp{\PP,\RR}$,
where initially $\PP = \DP(\RR)$.
To this end, many \emph{DP processors} have been proposed.
\NaTT implements the following DP processors:

\iffullversion
\def\myparagraph#1{\subsection{#1}}
\fi

\myparagraph{Dependency Graph Processor}
\REV{This}{The} processor decomposes
	\REV{a }{}DP problem $\Tp{\PP,\RR}$ into $\Tp{\PP_1,\RR} \dots \Tp{\PP_n,\RR}$
	where $\PP_1, \dots, \PP_n$ are the
	\emph{strongly connected components (SCCs)} of the
	\emph{dependency graph} \cite{HM04,GTS04}.
	Since the dependency graph is not computable in general,
	several approximations called
	\emph{estimated dependency graphs (EDGs)} have been proposed.
	\NaTT implements the EDG proposed in \cite{GTS05}.

\myparagraph{Reduction Pair Processor}
\REV{This}{The} processor forms the core of \NaTT.
	A \emph{reduction pair} is a pair $\Tp{\VGS,\VGT}$ of orders \st
	$\VGT$ is \emph{compatible} with $\VGS$ 
	(\ie, ${\VGS}\cdot{\VGT}\cdot{\VGS} \subseteq {\VGT}$),
	both of $\VGS$ and $\VGT$ are stable under substitution,
	$\VGS$ is monotone and $\VGT$ is well-founded.
	From a DP problem $\Tp{\PP,\RR}$,
	if all the involved rules are weakly decreasing
	(\ie, $\PP \cup \RR \subseteq {\VGS}$),
	strictly decreasing rules in $\PP$ (\wrt $\VGT$) can be removed.
	A great number of techniques for obtaining reduction pairs \REV{have been}{are} proposed so far.
	\NaTT supports the following\REV{ ones}{}:
\begin{itemize}
\item
	Some \emph{simplification orders} combined with \emph{argument filters} \cite{AG00}:
\vspace{2pt}
\begin{itemize}
\item
	the \emph{Knuth-Bendix order (KBO)} \cite{KB70} and its variants including
	KBO with \emph{status} \cite{S89},
	the \emph{generalized KBO} \cite{MZ97} and
	the \emph{transfinite KBO} \cite{LW07,WZM12},
\vspace{2pt}
\item
	the \emph{recursive path order} \cite{D82}
	and the \emph{lexicographic path order (LPO)} \cite{KL80},
\end{itemize}
\vspace{4pt}
\item
	\emph{polynomial interpretations (POLO)} \cite{L79,AG00}
	and its variants,
	including certain forms%
\footnote{Here, negative values are allowed only for \REV{the }{}constant part.} of
	\emph{POLO} with negative constants \cite{HM04b} and
	\emph{max-POLO} \cite{FGMSTZ08},
\vspace{2pt}
\item the \emph{matrix interpretation method}
\iffullversion
	\cite{HW06,EWZ08},
\fi
and
\vspace{2pt}
\item the \emph{weighted path order (WPO)} \cite{YKS13,YKS14}.
\end{itemize}
\REV{%
	Note that all of the above mentioned reduction pairs are subsumed by WPO.
	That is, by implementing WPO we obtain the other reduction pairs for free.
}{%
	While \NaTT supports such many techniques, all
	the reduction pairs obtained by these techniques
	are obtained just by implementing WPO,
	which subsumes them as instances.
}%
We discuss the implementation 
\REV{details}{in detail} in Section~\ref{sec:reduction}.

\myparagraph{Rule Removal Processor}
	In the worst case, the size of dependency pairs is
\REV{quadratic in}{proportional to square of}
	the size of the input TRS $\RR$.
	Hence it is preferable to reduce the size of $\RR$ before
	computing dependency pairs.
	To this end \NaTT applies the \emph{rule removal processor} \cite{GTS04}.
	If all rules in $\RR$ are weakly decreasing \wrt
	a \emph{monotone} reduction pair,
	then the processor removes strictly decreasing rules from $\RR$.
	The required monotonicity of a reduction pair is obtained by choosing
	appropriate parameters for the implementation of WPO described above.

\myparagraph{Uncurrying Processor}
	Use of uncurrying for proving termination is proposed for
	\emph{applicative} rewrite systems in \cite{HMZ13}.
	The uncurrying implemented in \NaTT is similar to the
	generalized version proposed in \cite{ST11},
	in the sense that it does not assume \emph{application symbols} to be binary.
	A symbol $f$ is considered as an application symbol if all the following
	conditions hold:
	\begin{itemize}
	\item $f$ is defined and has positive arity,
	\item
		a subterm of the form $f(x,\dots)$ does not occur
		in any left-hand-sides of $\RR$,
	\item
		a subterm of the form $f(g(\dots),\dots)$ occurs
		in some right-hand-side of $\RR$.
	\end{itemize}
	If such application symbols are found, then $\RR$ is uncurried \wrt
	the uncurrying TRS $\UU$ that consists of the following rules:%
\footnote{%
	The notation is derived from the \emph{freezing} technique \cite{X98}.
}
	\[
		f(f^lg(x_1,\dots,x_m),y_1,\dots,y_n) \to f^{l+1}g(x_1,\dots,x_m,y_1,\dots,y_n)
	\]
	for every $g \neq f$ and $l$ less than the \emph{applicative arity}%
\footnote{%
	Applicative arities are taken 
	so that \emph{$\eta$-saturation} is not needed.
} of $g$%
\REV{,
	where $f^0g$ denotes $g$ and $f^{l+1}g$ is a new function symbol of arity $m+n$}{}.

\section{The Weighted Path Order}
\label{sec:reduction}

As we mentioned in the introduction, \NaTT
implements only WPO for obtaining reduction pairs.
WPO is parameterized by
(1) a \emph{weight algebra} which specifies how weights are computed,
(2) \REV{a \emph{precedence} on}{\emph{precedences} of} function symbols, and
(3) \REV{a \emph{status function}}{\emph{statuses}} which
\REV{specifies}{specify} how arguments are compared.
In the following sections, we present some options which
\NaTT provides for specifying search spaces for these parameters.

\subsection{Templates for Weight Algebras}

One of the most important tasks
in proving termination by WPO is finding an appropriate weight algebra.
In order to reduce the task to an SMT problem,
\NaTT considers \emph{template algebras}
\REV{over integers}{on integer domain}.
Currently the following template algebras are implemented:

\begin{itemize}
\item
	The algebra $\Apol$ indicates that
	weights of terms are computed by a linear polynomial.
	Interpretations are in the following shape:
	\begin{equation}\label{eq:sum}
		f_\Apol(x_1,\dots,x_n) =
		w_{f} + \sum_{i=1}^n{c_{f,i} \cdot x_i}
	\end{equation}
	where
	the \emph{template variables}
	$w_f$ and $c_{f,1}, \dots, c_{f,n}$
	should be decided by an external SMT solver.
\smallskip
\item
	The algebra $\Amax$ indicates that
	weights are computed using the $\max$ operator.
	A symbol $f$ with arity $\ge 1$ is interpreted in the following shape:
	\begin{equation}\label{eq:max}
		f_\Amax(x_1,\dots,x_n) = 
		\max_{i=1}^{n} (p_{f,i} + c_{f,i} \cdot x_i)
	\end{equation}
	where $p_{f,1},\dots,p_{f,n}$ are template variables.
	For constant symbols,
	interpretations of the shape \eqref{eq:sum} are used.
	Since the operator $\max$ is not usually supported by SMT solvers,
	these interpretations are encoded \REV{as}{to} quantifier-free formula\REV{s}{} using
	the technique presented in \cite{YKS13}.
\smallskip
\item
	The algebra $\Amp$ combines both form\REV{s}{} of interpretations described above.
	Since it is inefficient to consider all combination\REV{s}{} of these interpretations,
	$\Amp$ decides the shape of interpretations according to the following
	intuition:
	If a constraint such as $f(x) > g(x,x)$ appears, then
	$g$ is interpreted as $g_\Amax$,
	because the imposed constraint
	$c_{f,1} \ge c_{g,1} \And c_{f,1} \ge c_{g,2}$
	is easier than
	$c_{f,1} \ge c_{g,1} + c_{g,2}$, which would be imposed
	by \REV{the interpretation }{}$g_\Apol$\REV{}{ interpretation}.
\end{itemize}

The template variables introduced above are partitioned into two groups:
\REV{}{the }template variables $w_f, p_{f,1},\dots,p_{f,n}$ are grouped in the \emph{constant part},
and template variables $c_{f,1},\dots,c_{f,n}$ are in the \emph{coefficient part}.
For efficiency,
it is important to properly restrict the range of these variables.

\subsection{Classes of Precedences}

\NaTT offers ``quasi'' and ``strict'' precedences, as well as
an option to disable them (\ie,
all symbols are considered to have the same precedence).
For reduction pairs using precedences,
\REV{%
	we recommend quasi-precedences which are chosen by default,
}{%
	quasi-precedences are recommended and supposed by default,
}%
as the encoding follows the technique of \cite{ZHM09}
that naturally encodes quasi-precedences.

\subsection{Classes of \REV{Status Functions}{Statuses}}

\NaTT offers three classes of \REV{\emph{status functions}}{\emph{statuses}}:
``total'',
``partial'' and ``empty'' \REV{ones}{statuses}.
The standard notions of \REV{\emph{status functions}}{statuses} are total ones that
were introduced to admit
\emph{permutation} of arguments when comparing them 
lexicographically from left to right (\cf \cite{S89}).
Such a comparison appears in many well-known reduction pairs;
famous examples are LPO and KBO.
By combining the idea of argument filters,
\REV{status functions}{statuses}
\REV{have recently been}{are recently} generalized to \emph{partial} ones,
that do not only \emph{permute} but may also drop some arguments \cite{YKS13b}.
\REV{A partial status is}{Partial statuses are}
beneficial for KBO,
and even more significant when combined with WPO \cite{YKS14}.
The extreme case of a partial status is the ``empty'' status,
that drops all arguments and so no comparison of arguments will be
performed. 
This option corresponds to the nature of interpretation methods, \eg POLO,
if precedences are also disabled.

\subsection{Obtaining Well-known Reduction Pairs}

Although most of the existing reduction pairs are subsumed by WPO,
some of them are still useful for improving efficiency,
due to the restricted search space and simplified SMT encoding.
We list parameters that correspond\REV{}{s} to some known reduction pairs in
Tables~\ref{tab:parameters} and \ref{tab:parameters 2}.
Note here that the effects of non-collapsing argument filters
are simulated by allowing $0$-coefficients in the weight algebra.
Thus \NaTT has a dedicated
implementation only for \emph{collapsing} argument filters, and
implementations of usable rules for
interpretation methods and path orders are smoothly unified.
\footnotetextlabel{KBO}{%
	Further constraints for \REV{}{the }\emph{admissibility} are imposed.}
\begin{table}[tb]
\caption{\label{tab:parameters}%
Parameters for some monotone reduction pairs.}%
\centering
\begin{tabular}{l@{\ }|cccccl}
	Technique	&template&coefficient&constant&precedence&status
\\	\hline
	Linear POLO	&$\Apol$&$\PosInt$	&$\Nat$	&no	&empty
\\	LPO			&$\Amax$&$\{1\}$	&$\{0\}$&yes&total
\\	KBO\footnoteref{KBO}
				&$\Apol$&$\{1\}$	&$\Nat$	&yes&total
\\	Transfinite KBO\footnoteref{KBO}
				&$\Apol$&$\PosInt$	&$\Nat$	&yes&total
\end{tabular}
\end{table}

\section{Cooperation with SMT Solvers}
\label{sec:SMT}

\NaTT is designed to work with any \SMTLIB 2.0 compliant solvers
that support at least \texttt{QF\_LIA} logic,
for which various efficient solvers exist\REV{}{s}.%
\footnote{\Cf the Satisfiability Modulo Theories Competition, \url{http://smtcomp.org/}.}
\NaTT extensively uses SMT encoding techniques 
for finding appropriate reduction pairs;
the conditions of reduction pair processors are
encoded into the following SMT constraint:
\begin{equation}
\label{eq:pair}
	\BigAnd_{l \to r \in \RR} \Encode{l \VGS r} \And
	\BigAnd_{s \to t \in \PP} \Encode{s \VGS t} \And
	\BigOr_{s \to t \in \PP} \Encode{s \VGT t}
\end{equation}
where each $\Encode{l \VGSopt r}$ is an SMT formula
that represents the condition $l \VGSopt r$.
In the remainder of this section,
we present two techniques for handling such constraints that
contribute to the efficiency of \NaTT.

\begin{table}[tb]
\caption{\label{tab:parameters 2}%
Parameters for some (non-monotone) reduction pairs.}%
\centering
\begin{tabular}{l@{\ }|cccccl}
	Technique	&template&coefficient&constant&precedence&status
\\	\hline
	Linear POLO
				&$\Apol$&$\Nat$		&$\Nat$	&no	&empty
\\	Max-POLO
				&$\Amp$	&$\Nat$		&$\Int$	&no	&empty
\\	LPO + argument filter
				&$\Amax$&$\{0,1\}$	&$\{0\}$&yes&total
\\	KBO + argument filter
				&$\Apol$&$\{0,1\}$	&$\Nat$	&yes&total
\\	Matrix interpretations
				&$\Apol$&$\Nat^{d\times d}$&$\Nat^d$&no	&empty
\\	WPO($\Ams$)
				&$\Amp$	&$\{0,1\}$	&$\Nat$	&yes&partial
\end{tabular}
\end{table}
\subsection{Use of Interactive Features of SMT Solvers}
\label{sec:interactive}

In a typical run of termination verification,
constraints of the form \eqref{eq:pair} are generated and solved many times,
and each encoding sometimes involves thousands of lines of SMT queries
with a number of template and \REV{auxiliary}{temporal} variables.
Hence runtime spent for the SMT solver forms a large part of
the overall runtime of the tool execution.
\NaTT tries to reduce the runtime by using
\emph{interactive} feature\REV{s}{} of SMT solvers,%
\footnote{%
\REV{%
	\NaTT is not the first tool to use the interactive features of SMT solvers.
	For example, the Houdini implementation in Boogie uses the features \cite{LQL12}.
}{}}
which \REV{are}{is} specified in \SMTLIB 2.0.

For each technique of reduction pairs,
the encoded formula of the constraint
$\BigAnd_{l \to r \in \RR} \Encode{l \VGS r}$ need not be changed
during a run, as far as $\RR$ is not modified.%
\footnote{%
	Although rules in $\RR$ may be removed by considering \emph{usable rules},
	the formula still need not be changed, since
	it can be simulated by negating a propositional variable that
	represents whether the rule is usable or not.
}
Hence, when a reduction pair processor is applied for the first time,
the back-end SMT solver is initialized according to the following pseudo-script:
\begin{Script}
	(assert ($\BigAnd_{l \to r \in \RR} \bigl(u_{l \to r} \Then \Encode{l \VGS r}\bigr)$))\\
	(push)
\end{Script}%
where $u_{l \to r}$ is a boolean variable denoting
whether the rule $l \to r$ is usable or not.
When the processor is applied to an SCC $\PP$,
the following script is used:
\begin{Script}
	(assert ($\BigAnd_{s \to t \in \PP} \Encode{s \VGS t} \And
		\BigOr_{s \to t \in \PP} \Encode{s \VGT t}$))\\
	(check-sat)
\end{Script}
Then, if a solution is found by the SMT solver,
\NaTT analyzes the solution using \REV{the }{}\texttt{get-value} command.
After this analysis, the command
\begin{Script}
 	(pop)
\end{Script}
is issued to clear the constraints due to $\PP$ and go back to
the context saved by the \texttt{(push)} command.
In order to derive the best performance of the solver,
\begin{Script}
	(reset)
\end{Script}
is also issued in case sufficiently many rules
become unusable (\eg, 1/3 of \REV{the }{}rules in $\RR$) from $\PP$.
All these commands, \texttt{push}, \texttt{pop} and \texttt{reset}
are expected to be available in SMT-LIB 2.0 compliant solvers.

\subsection{Use of Linear Arithmetic}
\label{sec:linearization}

Note that expressions of \REV{the }{}form \eqref{eq:sum} or \eqref{eq:max} are nonlinear,
due to the coefficients $c_{f,1},\dots,c_{f,n}$.
However, not many SMT solvers support \emph{nonlinear} arithmetic,
and even if they do, they are much less scalable than they are for linear arithmetic.
Hence, we consider reducing the formulas to linear ones
by restricting the range of $c_{f,1},\dots,c_{f,n}$ \eg to $\{0,1\}$.
Although the idea is inspired by \cite{BLNRR09},
\NaTT uses a more straightforward reduction using
\texttt{ite} (\emph{if-then-else}) expressions.
Each coefficient $c_{f,i}$ is replaced by
the expression \texttt{(ite $b_{f,i}$ 1 0)}
where $b_{f,i}$ is a propositional variable,
and then multiplications are reduced according to the rule:
\[
	\texttt{(* (ite $e_1$ $e_2$ $e_3$) $e_4$)}\ \to\ 
	\texttt{(ite $e_1$ (* $e_2$ $e_4$) (* $e_3$ $e_4$))}
\]
It is easy to see that this reduction terminates and
linearizes expressions of \REV{the }{}form \eqref{eq:sum} or \eqref{eq:max}.
It is also possible to avoid \REV{an explosion of the size}{explosion in sizes} of formulas
by introducing a \REV{auxiliary}{temporal} variable for the \REV{duplicated}{duplicating} expression $e_4$.

\def\ff{\mathsf{f}}
\def\aa{\mathsf{a}}
\def\bb{\mathsf{b}}
\begin{example}
\def\ccff{c_{\kern.1em\ff,1}}
\def\bbff{b_{\kern.1em\ff,1}}
	Consider the constraint $\ff(\ff(\aa)) > \bb$
	interpreted in the algebra $\Apol$, and
	suppose that the range of $\ccff$ is restricted to $\{1,2\}$.
	The interpretation of the term $\ff(\ff(\aa))$ is
	reduced as follows (written as S-expressions):
	\begin{align*}
		\Encode{\ff(\ff(\aa))}\ &=\ 
		\texttt{(+ $w_\ff$ (* (ite $\bbff$ 2 1) $\Encode{\ff(\aa)}$))}
	\\
		&\to\ 
		\texttt{(+ $w_\ff$ (ite $\bbff$ (* 2 $\Encode{\ff(\aa)}$) $\Encode{\ff(\aa)}$))}
	\end{align*}
	Similarly, for $\ff(\aa)$ we obtain
	\[
		\Encode{\ff(\aa)}\ \to\ 
		\texttt{(+ $w_\ff$ (ite $\bbff$ (* 2 $w_\aa$) $w_\aa$))}
	\]
	Now, the constraint $\Encode{\ff(\ff(\aa)) > \bb}$ is expressed by the
	following script:
	\begin{Script}
		(define-fun $v$ (+ $w_\ff$ (ite $\bbff$ (* 2 $w_\aa$) $w_\aa$)))\\
		(assert (> (+ $w_\ff$ (ite $\bbff$ (* 2 $v$) $v$) $w_\bb$)))
	\end{Script}
\end{example}

In contrast to SAT encoding techniques \cite{FGMSTZ07,FGMSTZ08,EWZ08},
we do not have to care about the bit-width for \REV{the }{}constant part and
intermediate results.
It is also possible to indicate \REV{that \NaTT should}{\NaTT to} keep formulas nonlinear,
and solve them using SMT solvers that support \REV{\texttt{QF\_NIA}}{\texttt{QF\_NRA}} logic.
Our experiments on TPDB%
\footnote{The Termination Problem Data Base,
\url{http://termination-portal.org/wiki/TPDB}.}
problems, however, suggests
that use of nonlinear SMT solving is impractical for our purpose.

\cut{
\subsection{Strategy for Choosing SCCs}

After applying the dependency graph processor,
dependency pairs are split into SCCs that can be handled independently.
Nonetheless,
the order of SCCs handled by a termination tool affects
the average performance of the tool.
This fact is explained as follows:
While all dependency pairs must be removed to conclude termination of a TRS,
a tool can immediately conclude \emph{nontermination}
(\ie, exit with ``\textsf{no}'') if it finds
a single SCC that admits an infinite chain.
Moreover, since finiteness of an SCC is an undecidable problem,
it is always possible that there exists an SCC for which
all the processors implemented in the tool fail to apply.
In this case it is nonsense to checking finiteness of the other SCCs,
and hence the tool can immediately \emph{give up}
(\ie, exit with ``\textsf{maybe}'').
Thus in general, it is more efficient to first handle SCCs
that are easier to decide if DP processors can be applied or not.

To estimate this easiness,
\NaTT considers the number of nodes in the SCC.
Hence smaller SCCs are processed first, and larger ones later.
This strategy can be elegantly implemented as follows:
\begin{enumerate}
\item
	Compute the initial set of SCCs in $\EDG(\RR)$,
	and push them in a stack according to the ascending order of the size.
\item
	Pop an SCC $\PP$ from top of the stack, which is the smallest one.
	Then try applying available processors on $\PP$ to obtain a simpler version $\PP'$.
	If this phase fails, try finding a loop or return \textsf{maybe}.
\item
	Decompose $\PP'$ into SCCs.
	Since $\PP'$ is smaller than any SCCs in the stack,
	so are the new SCCs.
	Hence just push them on top of the stack according to the order of the size.
\end{enumerate}
}

\section{Design}
\label{sec:design}

The source code of \NaTT consists of about 6000 lines of code\REV{}{s}
written in OCaml.\footnote{\url{http://caml.inria.fr/}}
About 23\% \REV{is consumed by}{are consumed for} interfacing SMT solvers,
where some optimizations for encodings are also implemented.
Another 17\% \REV{is}{are} for parsing command-lines and TRS files.
The most important part of the source code is
the 40\% devoted \REV{to}{for}
the implementation of WPO, the unified reduction pair processor\REV{}{s}.
Each of the other processors implemented \REV{}{merely }consumes less than 3\%.
For computing SCCs, the third-party library \texttt{ocamlgraph}%
\footnote{\url{http://ocamlgraph.lri.fr/}} is used.

\subsection{Command Line Interface}\label{sec:command}

The command line of \NaTT \REV{has}{is in} the following syntax:
\begin{center}
\tt
	./NaTT [FILE] [OPTION]... [PROCESSOR]...
\end{center}
To execute \NaTT, an \SMTLIB 2.0 compliant solver must be installed.
By default, \Zthree version 4.0 or later%
\footnote{\url{http://z3.codeplex.com/}}
is supposed to be installed in the path.
Users can specify other solvers by \REV{the }{}\texttt{--smt "COMMAND"} option,
where the solver invoked by \texttt{COMMAND}
should process \SMTLIB 2.0 scripts given \REV{on}{through} the standard input.

The TRS whose termination should be verified is read
from either the specified \texttt{FILE} or the standard input.%
\footnote{
The format is found at
\url{https://www.lri.fr/~marche/tpdb/format.html}.
}
Each \texttt{PROCESSOR} is either an order 
(\eg \texttt{POLO}, \texttt{KBO}, \texttt{WPO}, \etc, possibly followed by options), or
\REV{a }{}name of other processors (\texttt{UNCURRY}, \texttt{EDG}, or \texttt{LOOP}).
Orders preceding \REV{the }{}\texttt{EDG} processor should be
monotone reduction pairs and applied as rule removal processors
before computing the dependency pairs.
Orders following \REV{the }{}\texttt{EDG} processor are applied as
reduction pair processors to each SCC in the EDG.
A list of available \texttt{OPTION}s and \texttt{PROCESSOR}s
can be obtained via \texttt{NaTT --help}\REV{}{ command}.

\subsection{The Default Strategy}
\label{sec:default}
In case no \texttt{PROCESSOR} is specified,
the following default strategy will be applied:
\begin{itemize}
\item
	As a rule removal processor,
	POLO with coefficients in $\{1,2\}$ and constants in $\Nat$ is applied.
\item
	Then \REV{}{applies }the uncurrying processor\REV{ is applied}{}.
\item
	The following reduction pair processors are applied (in this order):
	\smallskip
	\begin{enumerate}
	\item POLO with coefficients in $\{0,1\}$ and constants in $\Nat$,
	\item algebra $\Amax$ with coefficients in $\{0,1\}$ and constants in $\Nat$,
	\item LPO with quasi-precedence\REV{}{s},
		status\REV{}{es} and argument filter\REV{}{s},
	\item algebra $\Amp$ with coefficients in $\{0,1\}$ and constants in $\Int$,
	\item WPO with
		quasi-precedence, partial status\REV{}{es},
		algebra $\Amp$, coefficients in $\{0,1\}$ and constants in $\Nat$,
	\item matrix interpretations
		with $\{0,1\}^{2\times2}$ matrices and $\Nat^2$ vectors.
	\end{enumerate}
	\smallskip
\item
	If all the above processors fail, then a (naive) loop detection is performed.
\end{itemize}

\section{Assessment}
\label{sec:assessment}
\iffullversion
\REV[10]{%

	In this section,
	we verify the significance of the contributions of \NaTT
	by experiments and
	by its result in the termination competition.

\subsection{Effects of Optimizations}
	First, we verify the effect of the optimizations proposed in \prettyref{sec:SMT}.
	The experiments are run on a server equipped with 
	a quad-core Intel Xeon E5-3407v2 processor running at a clock rate of 2.40GHz
	and 32GB of main memory.
	As the SMT solver, we choose \Zthreevar{4.3.2}.%

	In \prettyref{tab:effects}, we compare the following options of \NaTT.
	\begin{itemize}
	\item
		The `non-linear' row considers interpretations of the non-linear shape of
		\eqref{eq:sum} and \eqref{eq:max},
		and directly solves the encoded problem via \texttt{QF-NIA} logic.
		To achieve a practical runtime, the constant part is bounded by upper bound $3$.
	\item
		The `linearized' row applies
		the linearization proposed in \prettyref{sec:linearization}.
	\item
		The `interactive' row further uses the interactive features as
		proposed in \prettyref{sec:interactive}.
		This option is the default of \NaTT.
	\end{itemize}
}{%
}%
	\begin{table}[tb]
\REV[10]{%
		\caption{Effects of the optimizations.\label{tab:effects}}%
		\begin{center}
		\begin{tabular}{c@{\qquad}c@{\quad}c@{\quad}c@{\quad}c@{\quad}cc}
		\hline
			option		&yes&no	&maybe	&T.O.	&time\\
		\hline
			non-linear	&767&170&368	&158	&13138.11\\
			linearized	&848&173&429	&13		&2161.43\\
			interactive	&848&173&429	&13		&1865.50\\
		\hline
		\end{tabular}
		\end{center}
}{%
}%
	\end{table}
\REV[10]{%
	In the table,
	we observe a dramatic improvement by the linearization of
	\prettyref{sec:linearization}.
	The use of interactive features of SMT solvers
	may look less significant,
	but the runtime improves by almost 10\%.

	\subsection{Results in the Termination Competition}
}{%
}%
\fi
Many tools have been developed for proving termination of TRSs,
\REV{and}{so that} the international termination competition
\REV{has}{have} been held annually for a decade.
\NaTT participated 
in the \emph{TRS Standard} category of the full-run 2013,%
\iffullversion
	\footnote{\url{http://termcomp.uibk.ac.at/}}
\fi
where the other participants are versions of:
\AProVE,\footnote{\url{http://aprove.informatik.rwth-aachen.de/}}
\TTTT,\footnote{\url{http://cl-informatik.uibk.ac.at/software/ttt2/}}
\muterm,\footnote{\url{http://zenon.dsic.upv.es/muterm/}}
and \Wanda.\footnote{\url{http://wandahot.sourceforge.net/}}
Using the default strategy described in Section~\ref{sec:default},
\NaTT (dis)proves termination of 982 TRSs
\iffullversion
	(unfortunately, the competition version of \NaTT failed to input 36 problems
	due to a bug in parser)
\fi
and comes next to (the two versions of) \AProVE,
the constant champion of the category.
It should be noticed that \NaTT proved termination of 3\REV{4}{6} TRSs
out of the 159 whose termination could not be proved by any other tool\REV{}{s}.
\NaTT is notably faster than the other competitors;
it consumed only 21\% of \REV{the }{}time compared to \AProVE, the second fastest.
We expect that we can further improve efficiency by optimizing 
to multi-core architecture; currently, \NaTT runs in almost single thread.

\NaTT also participated in the \emph{SRS Standard} category.
However, the result is not as good as it is for TRSs.
This is due to the fact that
the default strategy of Section~\ref{sec:default} is designed only for 
non-unary signatures.
Indeed,
\REV{%
	when a unary symbol is considered,
	an interpretation of the form \prettyref{eq:max} is 
	equivalent to one of the form \prettyref{eq:sum},
}{%
	use of $\max$ makes sense only if it has at least two arguments.
}%
It should be improved by choosing \REV{a }{}strategy depending on \REV{the }{}shape of input TRSs.

\section{Conclusion}
\label{sec:conclusion}

We described the implementation and techniques of the termination tool \NaTT.
The novel implementation of the weighted path order is described in detail,
and some techniques for cooperating SMT solvers are presented.
Together with these efforts, \NaTT is one of the most efficient and 
strongest tools for proving termination of TRSs.
\REV[2]{%
\iffullversion

	Because of its efficiency, \NaTT is especially strong on larger systems.
	In general, a larger input TRS requires
	a larger proof script to be produced,
	which is quite difficult to be checked by hand.
	Thus our future work is to produce proofs in
	the \emph{certifiable proof format}.%
	\footnote{\url{http://cl-informatik.uibk.ac.at/software/cpf/}}
\fi
	\paragraph*{Acknowledgments}
	We thank the anonymous reviewers of this paper for careful inspections
	and constructive comments that improved the quality of this paper.
	We also thank Shaz Qadeer for information about Boogie.
	This work was supported by JSPS KAKENHI \#24500012.
}{%
}%
\bibliographystyle{plain}
\bibliography{../references}

\end{document}